# FP-GNN: a versatile deep learning architecture for enhanced molecular property prediction

Hanxuan Cai, Huimin Zhang, Duancheng Zhao, Jingxing Wu, Ling Wang[*]

Corresponding author: Ling Wang, Guangdong Provincial Key Laboratory of Fermentation and Enzyme Engineering, Joint International Research Laboratory of Synthetic Biology and Medicine, Guangdong Provincial Engineering and Technology Research Center of Biopharmaceuticals, School of Biology and Biological Engineering, South China University of Technology, Guangzhou 510006, China. E-mail: lingwang@scut.edu.cn

**Abstract**

Deep learning is an important method for molecular design and exhibits considerable ability to predict molecular properties, including physicochemical, bioactive, and ADME/T (absorption, distribution, metabolism, excretion, and toxicity) properties. In this study, we advanced a novel deep learning architecture, termed FP-GNN, which combined and simultaneously learned information from molecular graphs and fingerprints. To evaluate the FP-GNN model, we conducted experiments on 13 public datasets, an unbiased LIT-PCBA dataset, and 14 phenotypic screening datasets for breast cell lines. Extensive evaluation results showed that compared to advanced deep learning and conventional machine learning algorithms, the FP-GNN algorithm achieved state-of-the-art performance on these datasets. In addition, we analyzed the influence of different molecular fingerprints, and the effects of molecular graphs and molecular fingerprints on the performance of the FP-GNN model. Analysis of the anti-noise ability and interpretation ability also indicated that FP-GNN was competitive in real-world situations.

**Key words:** artificial intelligence; drug design and discovery; machine learning; molecular representation; graph attention networks

**Hanxuan Cai** is a graduate student at South China University of Technology. Her current research interests include machine learning, artificial intelligence-aided drug discovery (AIDD).

**Huimin Zhang** is a graduate student at South China University of Technology. Her research interests include artificial intelligence-aided drug discovery (AIDD).

**Duancheng Zhao** is a graduate student at South China University of Technology. His main research interests include deep learning for molecular generation.

**Jingxing Wu** is an undergraduate student at South China University of Technology. His research interests include machine learning and bioinformatics.

**Ling Wang** is an associate professor at South China University of Technology. He received a PhD from the School of Pharmaceutical Sciences at the Sun Yat-Sen University in 2014. His research focus on computer-aided drug design (CADD), artificial intelligence-aided drug discovery (AIDD) and medicinal chemistry.

**Introduction**

Predicting accurate molecular properties, including physicochemical and bioactive properties, as well as ADME/T (absorption, distribution, metabolism, excretion, and toxicity) properties, remains a fundamental challenge for molecular design, and especially for drug discovery and development. Quantitative Structure–Activity (Property) Relationship (QSAR/QSPR) modeling represents one of the most widely used and well-established computational approach for molecular properties prediction [1, 2]. QSAR/QSPR models are constructed using an empirical, linear, or nonlinear function that estimates an activity/property based on a chemical structure, followed by the application of those models to predict and design novel molecules with desired functional properties [3, 4]. With the continual accumulation of experimental data (e.g., chemical, biological and pharmacological related data), artificial intelligence (AI) and machine learning (ML) algorithms have become indispensable tools to establish QSAR/QSPR models that facilitate rapid, reliable, and affordable predictions and evaluations of physicochemical, biological, and ADME/T properties of small molecules in drug discovery practices.

Typically, ML-based QSAR/QSPR models have relied heavily on appropriate molecular representations [5]. Molecular representations play key roles in QSAR/QSPR analyses because of the large number of molecular features in such models, as well as the common requirement of model interpretability. Currently, molecular representations can be divided into three main categories, including molecular descriptors, molecular fingerprints, and molecular graphs. Molecular descriptors and fingerprints were derived from human expert domain knowledge for the comprehensive presentation of the constitutional, physicochemical, topological, and structural features of molecules [6-8]. Molecular descriptors and fingerprints are easily and quickly calculated, and can be used as inputs for both conventional ML (e.g., Naive Bayes (NB) [9], support vector machine (SVM) [10], random forest (RF) [11], extreme gradient boosting (XGBoost) [12]), and deep learning (e.g., deep neural networks) algorithms for QSAR/QSPR modeling tasks. However, molecular descriptor-based QSAR/QSPR models, and especially those for conventional ML, suffer from one major challenge in the era of big data: how to select the most important descriptors (called hand-crafted descriptors) related to a property of interest from a large number of predefined and calculable molecular descriptors [13]. This step is not only significant for the performance accuracy of the model, but is also directly related to the interpretability of the model. Recently, the emergence of deep learning (DL) approaches enables the elimination of tiresome expert and domain-wise feature constructions by delegating this task to a neural network that can extract the most valuable traits of the raw input data that are required to model the problem at hand [14, 15]. In contrast, for graph-based molecular representations, the atoms and bonds of a molecule are regarded as nodes and edges, and the aggregated node features are used by DL architectures, such as the graph convolutional network (GCN) [16], Graph Attention Network (GAT) [17], Attentive FP [18], Message Passing Neural Network (MPNN) [19], and Directed MPNN (D-MPNN) [20] for chemical learning tasks. Graph-based DL architectures have become popular and have been successfully employed in molecular property prediction tasks [21-26].

Although graph-based DL architectures are reported to yield state-of-the-art (SOTA) performance for molecular properties prediction tasks, whether graph-based DL models are better than conventional descriptors-based ML models for molecular properties learning tasks remains controversial. The majority of previous studies claimed that graph-based DL models were comparable or superior to conventional descriptors- or fingerprints-based ML models [20, 27, 28], while only few studies presented the opposite conclusion [29, 30]. For example, in 2021, Jiang et al. [30] performed comprehensive comparisons of graph-based DL models (i.e., GCN, GAT, MPNN, and Attentive FP that extracted features from molecular graphs) and conventional descriptor-based ML models (i.e., SVM, XGBoost, RF, and DNN) on 11 public datasets and demonstrated that the conventional descriptor-based models (especially RF and XGBoost ML algorithms) outperformed the graph-based DL models in terms of prediction accuracy and computational efficiency. Another recent study by Stepišnik and coworkers also reported a similar conclusion [31]. Currently, graph-based DL models still suffer from the potential limitation for insufficient modeling datasets, as it may be difficult for the automatic learning mechanism characteristics of graph neural networks (GNN) to learn robust graph representations from insufficient datasets [32]. In 2020, Rifaioglu et al. [28] discovered that graph- and fingerprint-based classifiers exhibited opposite trends when predicting the attributes of protein families. We hypothesize that the information captured by graphs or fingerprints are different, and may be complementary. Thus, the significant local chemical information contained in fingerprints may assist models to achieve superior results.

In this study, we introduce a new DL neural network architecture, FP-GNN (Figure 1), for molecular properties prediction. FP-GNN first operates over a hybrid molecular representation that combines molecular graphs and molecular fingerprints. It not only learns to characterize the local atomic environment by propagating node information from nearby nodes to more distant nodes using the attention mechanism in a task-specific encoding, but also provides a strong prior using fixed and complementary molecular fingerprints. We evaluated the FP-GNN model and other recently published graph-based DL algorithms (e.g., D-MPNN (Chemprop), Attentive FP, and HRGCN+) and a venerable fingerprint-based CML algorithm, XGBoost, against 13 commonly used public benchmarks. Compared to all baseline models, FP-GNN achieved comparable or superior performance in 11 out of 16 experiments on the 13 public datasets, thus, illustrating its strong out-of-the-box and SOTA performance when modeling a broad range of molecular properties.

FP-GNN was also tested on LIT-PCBA, which is an unbiased data set for ML and virtual screening (VS), and exhibited comparable or superior performance to fingerprint-based CML methods (e.g., NB, SVM, RF, and XGBoost) and graph-based DL methods (e.g., GCN and GAT). Additionally, FP-GNN performed well on 14 phenotypic screening datasets for breast cell lines compared to the XGBoost CML method and graph-based DL methods (e.g., GAT, GCN, MPNN and Attentive FP). All such results confirmed our hypothesis that molecular fingerprints can improve the generalization ability of graph-based DL algorithms. Anti-noise ability testing of FP-GNN also revealed its superiority over the Attentive FP, XGBoost, and HRGCN+ models, while maintaining high predictive power. Moreover, the interpretability of FP-GNN can deduce significant fragments from graph-based representations and important substructures from fingerprint-based representations, which can assist chemists in designing better molecules with desired functions or properties.

**Methods and materials**

**Graph neural networks with attention mechanism**

Molecules are natural graph-structured data and we therefore choose the spatial-GNN [33] to compute the information from molecular graphs. Before the data were inputted into the GNN model, we transformed each molecule into an undirected graph $G(V, E)$, where $V = \{x_1, x_2, ..., x_n\}$ is the node set representing atoms and $E$ is the edge set representing chemical bonds. The spatial-GNN updates each node by aggregating the information of itself and neighbors according to

$$h_i' = Aggregate_1(h_i, \sum_{j \in N(i)} h_j) \quad (1)$$

where $h_i$ is the vector of node i and $N(i)$ means the neighbors of i.

Finally, the model aggregates the total graph to the output according to

$$H = Aggregate_2(\sum_{i \in G} h_i') \quad (2)$$

As shown in Figure 1a, we used the attention mechanism [17] to update the node message. The graph attention mechanism pays attention to the effect of neighbors and computes the attention from node j to node i according to

$$e_{ij} = LeakyRelu(a \cdot [W_1 h_i || W_1 h_j]) \quad (3)$$

where $h_i \in R^l$, $W_1 \in R^{l' \times l}$, $a \in R^{1 \times 2l'}$ and $||$ indicates the concatenation operation. The attentions of all neighbors were then normalized according to

$$\alpha_{ij} = softmax(e_{ij}) = \frac{\exp(e_{ij})}{\sum_{k \in N(j)} \exp(e_{ik})} \qquad (4)$$

Attentions were used as weights to update node i as follow:

$$h'_i = elu(\sum_{j \in N(i)} \alpha_{ij} W_1 h_j) \qquad (5)$$

where $h'_i \in R^{l'}$. We then computed attentions many times and calculated the mean as the final attention:

$$h'_i = elu(\frac{1}{K} \sum_{k=1}^{K} \sum_{j \in N(i)} \alpha_{ij}^k W_1^k h_j) \qquad (6)$$

After updating all nodes, the output of the complete molecular graph was the mean of them:

$$H = (\frac{1}{N} \sum_{i=1}^{N} h'_i) \qquad (7)$$

**Initial molecule featurization**

Similar to other graph-based methods [18, 20], we used the properties of molecules to initialize the nodes of the molecular graphs before the data were imported into the GNN model. The atomic features are given in Supplementary Table 1.

**Molecular fingerprints**

Molecular fingerprints, which are bit strings mapped from a molecule according to different established rules, are abstract molecular representations. Molecular fingerprints are roughly divided into substructure key-based fingerprints, topological or path-based fingerprints, and circular fingerprints [34]. Three complementary fingerprints (MACCS fingerprint [35] , Pharmacophore ErG fingerprint [36], and PubChem fingerprint [37] were used in the FP-GNN model because they can complement and holographically express molecular characteristics [38]. The three fingerprints are described as follows:

MACCS fingerprint: a substructure key-based fingerprint using SMARTS pattern. MACCS contains most atomic properties, bond properties and atomic neighborhoods at diverse topological separations, which are meaningful for drug discovery. We choose the short variant of 1+166 bits for this study.

PubChem fingerprint: a substructure key-based fingerprint of 881 bits with extensive coverage of chemical structures.

Pharmacophore ErG fingerprint: a 2D pharmacophore fingerprint using an extended reduce graph (ErG) method and applying pharmacophore-type node descriptions to encode molecular properties.

**FP-GNN network architecture**

As shown in Figure 1b, the FP-GNN architecture first combined the molecular graph and three complementary molecular fingerprints into the flexible and dynamic neural network. The simplified molecular input line entry system (SMILES) notation of the molecule was inputted to two paths of the FP-GNN architecture.

On one path, three complementary fingerprints (MACCS, PubChem, and Pharmacophore ErG fingerprints), termed the mixed fingerprints, were concatenated according to

$$FP = (FP_{PubChem} \parallel FP_{MACCS} \parallel FP_{\text{Pharmacophore ErG}}) \quad (8)$$

The fingerprints vector was inputted into the artificial neural network (ANN) to obtain the following representation (Equation 9):

$$V' = W_2 \cdot FP + b \quad (9)$$

On the other path, the GNN model was used to capture the information of the molecular graph. The node representation was aggregated from itself and its neighbors by the attention mechanism. Finally, the average of all nodes was produced as the output to represent the molecular graph.

The outcomes received from the two paths were then fitted together and imported into fully connected layers to produce the ultimate output.

**Hyperparameter optimization and training protocol**

Similar to other DL models, the correct selection of hyperparameters can optimize the performance of the FP-GNN model. In the present study, the Hyperopt Python package [39] was employed to conduct Bayesian optimization of hyperparameters. Six hyperparameters were chosen: the dropout rate of GNN, the number of multi-head attentions, the hidden size of attentions, the hidden size and dropout rate of the fingerprint networks (FPN), and the ratio of GNN in FP-GNN.

FP-GNN was developed by the Pytorch framework. All FP-GNN models were trained on the SCUTGrid (SCUT supercomputing platform), which uses Matrox MGA G200e. The source codes for FP-GNN, as well as the training details are freely available on GitHub (https://github.com/idrugLab/FP-GNN).

**Benchmark datasets and performance evaluation metric**

The performances of the FP-GNN models were extensively evaluated using three benchmark datasets. First, 13 commonly used public datasets (Supplementary Table 2) relevant to drug discovery were used to test the performance of FP-GNN, including three physicochemical datasets (ESOL [40], FreeSolv [41] and Lipophilicity [42]), six bioactivity and biophysics datasets (MUV [43], HIV [44], BACE [45], PDBbind-C, PDBbind-R and PDBbind-F [46]), and four physiology and toxicity datasets (BBBP [47], Tox21 [48], SIDER [49] and ClinTox [50, 51]). Second, LIT-PCBA [52], a recently developed unbiased and realistic dataset that consists of 15 targets and 7844 confirmed active and 407,381 confirmed inactive compounds (Supplementary Table 4), was used to evaluate the performance of FP-GNN. Finally, 14 phenotypic screening datasets for breast cell lines (Table 2) were also employed to assess the predictive power of FP-GNN [53].

In accordance with the evaluation methods of the CML and DL models, the regression tasks were evaluated by root-mean-square error (RMSE), while the classification tasks were evaluated by the area under the receiver operating characteristic curve (ROC-AUC) or the area under the precision recall curve (PRC-AUC).

**Results and discussion**

**Performance of the FP-GNN network architecture on the public benchmark**

**datasets**

The 13 commonly used public benchmark datasets related to drug discovery from Wu et al. [27] were used to evaluate the predictive power of the FP-GNN models. As shown in Supplementary Table 2, the benchmark datasets encompassed three categories: physicochemical, bioactivity and biophysics, and physiology and toxicity. The sizes of the datasets varied widely and included small datasets (e.g., PDBbind-C only contains 168 molecules) and large datasets (e.g., MUV datasets containing 17 learning tasks and consisting of 93,087 molecules). For multi-task datasets, the average performance metric of each model was calculated to represent the final performance. ROC-AUC was used as the evaluation metric for all classification tasks, except the classification models established on the MUV datasets. For a highly imbalanced dataset, PRC-AUC could better reflect the performance of classification models than ROC-AUC. Since the ratio of actives to inactives in the MUV datasets was highly imbalanced, the PRC-AUC was used to evaluate the performance of the classification models based on the MUV datasets. Regression models were evaluated using RMSE. To fairly compare the published performance of the SOTA graph-based DL models (MoleculeNet [27], D-MPNN (Chemprop) [20], Attentive FP [18] and HRGCN+ [54]) and the advanced descriptors-based XGBoost [12] models on the public datasets, the same data-split code was adopted to randomly split each dataset into the training set, the validation set, and the test set, with a ratio of 8:1:1. In addition, BACE, BBBP and HIV datasets were split at the same ratio based on molecular scaffolds. To reduce the occasionality in data splitting and to ensure the reliability of the results, the hyperparameters were optimized.

The average value of the evaluation metrics of the FP-GNN models based on 10 different random seeds was computed as the final result.

The active compounds from the bioactivity and biophysics (Supplementary Table 2) datasets were measured based on their binding affinities for different biological targets. There is no doubt that accurately predicting the biological activities of small molecules for a given target can accelerate the discovery and development of new drug candidates. There was a total of eight learning tasks for this type of dataset (Supplementary Table 2), including four classification tasks based random- and scaffold-splitting methods for the HIV and BACE bioactivity datasets, one classification task based on the random-splitting method for the MUV bioactive dataset, and three regression tasks based random-splitting method for three biophysical datasets (PDBbind-C, PDBbind-R, and PDBbind-F). As shown in Table 1, FP-GNN performed best on three of the eight learning tasks, including the two classification learning tasks based on the scaffold-splitting of BACE and HIV, and one regression task of PDBbind-C. Chemprop achieved the four best performance tasks, including one classification task based on the random-splitting of BACE and HIV, and two regression tasks of PDBbind-F and PDBbind-R. Graph-based weave models from MoleculeNet performed best on the MUV dataset that contains 17 subtasks. Notably, FP-GNN achieved the second-best performance on the random-splitting of HIV, MUV, PDBbind-F and PDBbind-R. Although FP-GNN did not perform the best on some datasets, our model still performed comparatively well on those datasets.

Molecules from physiology and toxicity datasets record their effects will living bodies, such as the blood-brain barrier penetration (BBBP), the side effect resource (SIDER), and toxicities (Tox21 and ClinTox). Thus, those datasets are closely related to the physiology and toxicity properties of drugs. Precisely predicting the physiological and toxicological properties of compounds can rule out improper molecules in the early stages of drug discovery, which is beneficial for reducing the cost reduction of new drug development. However, it remains challenging to predict physiological and toxicological properties accurately. As shown in Table 1, FP-GNN achieved the three best classification performance results on the BBBP (from random- and scaffold-splitting methods) and SIDER datasets, while Chemprop performed best on Tox21 and XGBoost performed best on ClinTox. FP-GNN also exhibited better performance than the Weave models of MoleculeNet on the ClinTox dataset.

The physicochemical properties of a given drug can reflect its pharmacokinetic phases in the body. The physicochemical properties of molecules play a key role in the development of candidate drugs. Therefore, the accurate prediction of the physicochemical properties of molecules facilitates drug discovery and deveopment. FreeSolv, ESOL and Lipophilicity datasets were used to evaluate the predictive ability of the FP-GNN network architecture for physicochemical properties. Table 1 illustrates that FP-GNN performed best on the FreeSolv dataset, HRGCN+ performed best on the ESOL dataset, and Attentive FP performed best on the Lipophilicity dataset. Although FP-GNN performed worse than Attentive FP on the Lipophilicity dataset, it outperformed the other graph-based DL methods (e.g., GCN, MPNN and Weave) in MoleculeNet.

The ultimate goal of building molecular property prediction models is to predict the properties of new molecules with novel scaffolds, to make them fall within the appropriate ranges of the desired properties. Consequently, the scaffold-based splitting method was used on the BACE, BBBP and HIV datasets to ensure that the scaffolds in the training set, validation set, and test set were as distinct as possible. As shown in Table 1, the performance of scaffold-splitting classification models was lower than that of models based on random-splitting. Those data suggest that the scaffold-based splitting method was more challenging for learning tasks. Our FP-GNN models performed best on all three datasets using scaffold-based splitting and showed the same outstanding performance as the random-splitting method. All such results demonstrate that FP-GNN is stable in predicting molecules with new scaffolds.

Out of 16 learning tasks from 13 public benchmark datasets (Table 1), FP-GNN showed the best performance on seven tasks, while Chemprop exhibited the best performance on five tasks. MoleculeNet, Attentive FP, HRGCN+ and XGBoost performed best on one task each. Supplementary Table 3 summarizes how our FP-GNN compared to each of the baseline models, including four commonly used SOTA graph-based DL methods and the venerable descriptors-based ML method, XGBoost. Our FP-GNN model consistently matches or outperforms not only for each baseline individually (Supplementary Table 3), but also across all baselines (Table 1), indicating that coupling molecular graphs and fingerprints can improve the degree of generalization of graph-based DL algorithms to predict molecular properties better. The outstanding performance of FP-GNN on drug discovery-related datasets makes FP-GNN one of the most competitive DL methods in drug discovery practice.

**Performance of the FP-GNN network architecture on an unbiased and realistic LIT-PCBA dataset**

In 2020, Viet-Khoa Tran-Nguyen et al. [52] designed an unbiased and realistic dataset called LIT-PCBA, specifically dedicated to ML and VS methods. The LIT-PCBA dataset overcomes the obvious and hidden chemical biases of artificially-constructed public benchmark datasets (e.g., DUD, DUD-E and MUV) that are typically used by the community, and therefore, not overestimating of the true accuracy of ML methods. LIT-PCBA consists of 7844 confirmed active and 407,381 confirmed inactive compounds towards 15 targets, which were collected from the PubChem BioAssay (PCBA) dataset [55]. For each target, unbiased training and validation sets were constructed using the asymmetric validation embedding (AVE) method at the ratio of 3:1. The details of LIT-PCBA dataset are summarized in Supplementary Table 4. We therefore used this dataset to evaluate the predictive power of FP-GNN. Five fingerprint-based methods [56] (e.g., NB, SVM, RF, XGBoost and DNN) and two graph-based methods (e.g., GCN and GAT) were selected as the baseline models. All fingerprint-based models were constructed based on the Morgan fingerprint [57] and the mixed fingerprints (MACCS FP, PubChem FP and Pharmacophore ErG FP). According to the original paper and Jiang et al. [56], ROC-AUC was used to evaluate the performance of the classification models for the LITPCBA dataset.

As shown in Figure 2a, when compared to five Morgan fingerprint-based models and two graph-based models, FP-GNN exhibited the best performance on six targets (ADRB2, ALDH1, ESR1_ago, MAPK1, PPARG and TP53). Meanwhile, NB achieved the best performance on two targets (IDH1 and VDR), DNN achieved the best performance on two targets (FEN1 and OPRK1), GCN performed best on two targets (ESR1_ant and MTORC1), SVM, XGBoost and GAT achieved the best performance on one task each (PKM2, GBA and KAT2A, respectively). Compared to the mixed fingerprints-based models, FP-GNN also showed a similar outstanding performance (Figure 2b). The details of the direct comparisons between the FP-GNN model and each of the baseline models are listed in Supplementary Table 5. It is clear that not only did our FP-GNN models outperform the fingerprint-based models but also exhibited comparable or superior performance to the two classical graph-based DL models (GCN and GAT). Even on the most challenging LIT-PCBA dataset, FP-GNN also exhibited strong competitiveness and can be used to accurately predict the biological activity of molecules for drug discovery campaigns.

**Performance of FP-GNN compared to the SOAT graph-based and fingerprint-based models on cell-based phenotypic screening datasets**

Phenotypic-based screening (e.g., whole-cell activity), an original but indispensable drug screening method, has regained attention in recent years [58-62]. The phenotypic screening datasets for 13 breast cancer cell lines and one normal breast cell line were used to evaluate the performance of FP-GNN. Details of these datasets are shown in Table 2. Recently, He et al. [53] reported four graph-based DL models and one advanced fingerprint-based XGBoost model to predict the activities of molecules against those cell lines. Therefore, FP-GNN models were developed on the 14 cell-based phenotypic screening datasets and compared to the performances published for the five models. The same data split and evaluation metric (ROC-AUC) were adopted from He et al. The AUC value of the FP-GNN model for each cell line was calculated from the average AUC of the models based on 10 random seeds.

As shown in Table 2, FP-GNN performed best on eight out of 14 cell lines (i.e., MDA-MB-453, SK-BR-3, T-47D, MCF-7, BT-474, BT-20, BT-549 and HBL-100), while Attentive FP achieved the best performance on three cell lines (HS-578T, MDA-MB-231 and Bcap37), XGBoost achieved the best performance on two cell lines (MDA-MB-361 and MDA-MB-468), and GCN performed best on MDA-MB-435. Notably, our FP-GNN models were the second-best in HS-578T, MDA-MB-231 and MDA-MB-468 datasets. Importantly, FP-GNN achieved the overall best performance on these 14 cell lines, with the highest average AUC value of 0.849. FP-GNN exhibited excellent performance on the cell-based phenotypic screening datasets compared to the SOAT graph-based Attentive FP/GCN/GAT/MPNN models and one advanced fingerprint-based XGBoost model, thus, suggesting that FP-GNN holds great potential for phenotype-based drug discovery.

**The Ablation Experiment of FP-GNN.**

We investigated whether the information of local neighbors and global structures learned from the molecular graphs, and the chemical substructure information learned from the molecular fingerprints could complement each other and assist in optimizing our FP-GNN model.

To analyze the influence of the graph-based module and fingerprint-based module in the FP-GNN model, we counted the ratios of GNN in FP-GNN (Figure 3) based on the optimal set of hyperparameters for each of the 13 public datasets (Supplementary Table 6). As shown in Figure 3, more than half (54.3%) of the ratios of GNN in FP-GNN fell between 0.4 and 0.6, illustrating that the contributions of the two modules to the FP-GNN model were relatively balanced. In addition, pure GNN and pure FPN only accounted for approximately 4.3% of all models, demonstrating that coupling complementary molecular graph and fingerprint strategy to dynamic GNN can improve the performance of molecular property prediction.

The ablation experiment of FP-GNN was conducted on the unbiased and realistic LIT-PCBA dataset. Each model for the 15 targets was split into FPN and GNN models with the original hyperparameters. FP-GNN also used the same hyperparameters except that the ratio of GNN in FP-GNN modules was set to 0.5. As shown in Figure 4, FP-GNN models outperformed FPN and GNN in 10 out of 15 targets. FP-GNN models performed medium, with slightly lower performance than GNN, but evidently higher than FPN models on most of the other five targets (ESR1_ago, FEN1, KAT2A, MTORC1, and OPRK1). These results illustrated that FP-GNN combines the advantages of FPN and GNN to capture the complementary information of molecular graphs and fingerprints to achieve better performance. A possible reason for this finding was that we used the default parameter 0.5 as the ratio of GNN in FP-GNN modules when building the FP-GNN model, and less information was captured from the unfavorable GNN or FPN module to influence the performance of FP-GNN on the five targets. Collectively, combing molecular graphs and fingerprints can obtain local neighbor and complete structural information from the molecular graphs and substructures, as well as pharmacophore information from the molecular fingerprints, which will enable more accurate predictions of molecular properties.

**The influence of different types of fingerprints**

There are a variety of molecular fingerprints for different molecular representations, and the advantages of each are also distinct. Therefore, we explored the influence of different molecular fingerprints on the performance of our FP-GNN network architecture. The Morgan fingerprint is a circular fingerprint to record the structural environment features of each atom with a diameter of 4 and is the most common fingerprint used in QSAR/QSPR modeling [58, 63-67]. In addition to the mixture of three complementary fingerprints, we grafted the 1024-bits ECFP-4 fingerprint [68] (referred to as the Morgan fingerprint in the RDKit) into FP-GNN architecture and then tested it on public datasets.

As shown in Figure 5, FP-GNN models based on the mixed fingerprints performed better than the FP-GNN models based on the Morgan fingerprint, regardless of whether the analyses were performed on classification datasets (Figure 5a) or regression datasets (Figure 5b). In addition, we retrospect the performances of four CML methods (i.e., RF, SVM, NB, and XGBoost) and one DNN DL method using the mixed fingerprints and the Morgan fingerprint on the LIT-PCBA dataset. As shown in Figure 2, by counting the number of best-prediction models based on the mixed fingerprints and the number of best models based on the Morgan fingerprint, it was seen that the former did not exhibit absolute superiority (42 *vs.* 30, and the three results are equal). Furthermore, when comparing the performance of the Morgen fingerprint-based models and the mixed fingerprint-based models, there is a trend that simple NB and SVM methods can achieve more information from Morgan fingerprint, while advanced algorithms (RF, XGBoost and DNN) can capture more information from the mixed fingerprints. Meanwhile, the FP-GNN based on the mixed fingerprints exhibited better performance than FP-GNN based on the Morgen fingerprint (Figure 5). Thus, those data indicated that compared to the commonly used Morgen fingerprint, coupling the mixed fingerprints and molecular graph could achieve optimal complementarity to exhibit better performance.

The above-mentioned differences in complementarity may be related to the specific generation algorithms of fingerprints. The mixed fingerprints recorded most of the atomic and bond properties (MACCS fingerprint), extensive chemical structures and substructures (PubChem fingerprint) and pharmacophore feature (Pharmacophore ErG fingerprint) information, which may not be included in the features of molecular graphs. However, the Morgan fingerprint only records the local environmental information of atoms, which may be similar to the molecular graph features. Therefore, unlike the Morgen fingerprint, the mixed fingerprints can better complement the molecular graph features, and elicit better molecular representations.

**The Anti-noise Ability of FP-GNN**

DL models place extensive demands on data quality and generally require a large quantity of correct data. Obtaining sufficient high-quality data is still the central challenge in computer-assisted drug discovery [32]. Actually, the available data used in drug discovery practices are usually scarce and of mediocre quality. When the model is used in real-world scenes, the noises in the data will affect the training process and reduce the practicality of the model. Therefore, we ran FP-GNN on the noisy data to test its anti-noise ability.

We divided the HIV dataset (41,127 compounds) at a ratio of 8:1:1 to generate the training set, validation set, and test set. We ensured that the labels in the test set remained unchanged, while changing the labels in the training set and validation set according to a predetermined ratio to generate noise artificially. The anti-noise ability of FP-GNN was compared with two DL methods (Attentive FP and HRGCN+) and one advanced CML method (XGBoost) from Wu et al. [54]. The same data, data split, evaluation metric (ROC-AUC) and noise rates were also adopted from Wu et al. to ensure a fair comparison. Figure 6 indicated that FP-GNN achieved the SOTA performance in the anti-noise tests. Based on the excellent anti-noise ability of our FP-GNN model, it is foreseeable that it can handle poor data situations in real drug discovery scenarios.

**The Interpretation of FP-GNN**

In the field of drug discovery, an explicable model can assist in understanding the potential mechanism and capturing valuable information of molecules for specific tasks (e.g., lead optimization). FP-GNN has shown excellent performance on multiple benchmark datasets, driving us to explore the interpretability of the FP-GNN model.

The FP-GNN model developed based on the BBBP dataset that contains the blood-brain barrier permeability of molecules was used to analyze the interpretability of the model. Since the BBB can block most drugs and hormones, it is essential to accurately predict the BBB permeability of molecules for the development of drugs targeting central nervous system diseases. Facing the natural BBB that exists in the human body, hydrophobic molecules (low polarity and high ClogP) easily bypass the BBB, while the converse is true for hydrophilic molecules.

The FP-GNN architecture can compute the attentions of adjacent atoms and then map them to bonds connected to atoms (Figure 1). For a given molecule, the attention coefficient can be used to quantitatively characterize whether chemical fragments contribute more to the prediction of molecular properties. As shown in Figure 7, the portions of the molecule colored more darkly were more significantly in predicting whether the molecule can pass the BBB, while the role of the light-colored portions is less important. Considering an active molecule as an example (Figure 7a), most of the substructural groups of this compound are hydrophobic, laying the foundation for penetrating the BBB. The benzene ring (C7-C12, marked in red) of the molecule has the least polarity and maximum contribution to BBB penetration. We used ChemBioDraw (v.14.0.0.117) to further quantitatively analyze the ClogP values of these chemical fragments. Quantitative analyses of ClogP showed that the chemical portion marked in red had a lower polarity (ClogP = 2.142), while the grey mark portion had a higher polarity (ClogP = 1.389). In fact, our FP-GNN model paid great attention to the low-polarity benzene ring, which was also consistent with the prediction results as an active molecule. As shown in Figure 7b, for an inactive molecule, the dark portion (marked in red) represents an exposed substituent amino group that provides the majority of polarity to prevent the molecule from passing the BBB. The ClogP value of the chemical fragment in red is -0.905, while the ClogP of the fragment in grey is 0.934. The lower ClogP indicates that the red portion of the molecule was more hydrophilic and difficult to cross the BBB. The high attention marked in the red part from our FP-GNN model was consistent with the inactive prediction results. These cases not only demonstrate that our FP-GNN model was interpretable, but also hint the FP-GNN network architecture can learn the relationships between molecular substructures (chemical fragments) and their molecular properties. Thus, the prediction of highly favorable and unfavorable chemical fragments using the FP-GNN model will help design and optimize new molecules with desired properties or functions.

Besides the GNN module, we analyzed the interpretation of the FPN module. We chose the FreeSolv (Free Solvation) dataset that contains the hydration free energy of small molecules in water. The mixed fingerprints (MACCS FP, Pharmacophore ErG FP, and PubChem FP) that we used in the FPN model had 1489 bits in total. We changed the values of each bit in order and then inputted the mixed fingerprints to the training model. The effects made by different changing bits indicated the importance of fingerprints in the model. The more a modified value deviated from the original prediction, the more critical the fingerprint bit was in predicting the free solvation of molecules. The ten most significant bits are shown in Table 3. As shown in Table 3, substructures represented by the 4th, 5th, 7th, and 10th bits had strong polarity and high water-solubility, which play essential roles in the free solvation of molecules. We calculated the Pearson correlation coefficient between the hydration-free energy of molecules and these ten fingerprint bits. The Pearson values of the 3rd, 6th, and 10th bits were above 0.7, indicating that they exhibited a strong correlation. Thus, it can be seen that our model captured the significant part of fingerprints, and the prediction results from the FP-GNN model can be explained. As shown in Table 3, in the top ten crucial bits, there are four, three, and three bits coming from MACCS FP, Pharmacophore ErG FP and PubChem FP, respectively. Such results illustrate that three fingerprints together played an important role in the FP-GNN model.

**Conclusions**

In this study, we advanced a new DL architecture called FP-GNN, which first couples the graph attention network based on a molecular graph and the artificial neural network based on the mixed molecular fingerprints to generate more comprehensive molecular representations. The performance of FP-GNN on 13 classical public datasets revealed that our FP-GNN model performed outstandingly compared to four recently-published graph-based DL algorithms (MoleculeNet, Chemprop, Attentive FP, and HRGCN+) and the venerable XGBoost CML algorithm. We also evaluated the predictive power of FP-GNN on an unbiased and realistic LIT-PCBA dataset and 14 phenotypic drug screening datasets related to breast cancer cell lines. The evaluation outcomes further indicated that our FP-GNN model was highly competitive. Analyses of the influence of molecular graphs and fingerprints on the FP-GNN model, as well as the results of ablation experiments found that (1) molecular graphs and mixed molecular fingerprints in the FP-GNN architecture contributed to improve model prediction performance; and (2) embedding different fingerprints in the FP-GNN architecture affected its predictive performance. Currently, the mixture of three fingerprints consisting of atomic and bond properties, substructures, and pharmacophores can achieve optimal complementarity with the graph-based module. The ablation experiments also found that (3) complementary molecular representations by coupling molecular graphs and mixed molecular fingerprints played a key role in improving model performance. In addition, the excellent anti-noise ability of FP-GNN indicated that our FP-GNN model could solve noisy (poor) data in the natural scenes of drug discovery. Importantly, the FP-GNN model has intuitive interpretability and can identify important chemical fragments in a molecule, which can assist in designing and optimizing new molecules with desired properties or functions. Collectively, we expect that FP-GNN as a new DL architecture will help chemists, biologists and pharmacists predict molecular properties rapidly and effectively. There are several optimization routes for our future works. On the one hand, pre-trained methods may have great potential due to the insufficient quantity and poor quality of biological datasets. Extracting information from a large dataset before training the target dataset can ensure the prescribed minimum of prediction on the target dataset. On the other hand, when training a model on a specific dataset, it is feasible to import the information of the protein target into a modified FP-GNN model models and then combine features of molecules and the protein target to predict molecular properties collectively.

**Key points**

• We presented a DL architecture named FP-GNN to predict molecular properties, which couples the graph attention networks based on molecular graphs and the artificial neural networks based on the mixed molecular fingerprints to generate more comprehensive molecular representation.

• Extensive experimental results showed that FP-GNN was highly competitive compared to classic ML methods and state-of-the-art DL methods.

• The ablation experiments of FP-GNN indicated that information from molecular graphs and molecular fingerprints is complementary to improve the predictive power of molecular properties.

• The intuitive interpretability of the FP-GNN model can provide important chemical fragments to assist chemists and pharmacists in designing or optimizing new molecules with desired properties.

**Author contributions**

L.W. designed the study and L.W. and H.X.C. wrote the manuscript. H.X.C. performed the experiments and data analysis. H.M.Z, D.C.Z. and J.X.W. provided evaluation and suggestions. All authors contributed to the manuscript.

**Competing interests**

The authors declare no competing interests.

**Supplementary Data**

Supplementary data are available online at https://academic.oup.com/bib.

**Data and code availability**

The full datasets and the code for FP-GNN are available on GitHub at https://github.com/idrugLab/FP-GNN.

**Acknowledgements**

This work was supported in part by the National Natural Science Foundation of China (81973241) and the Natural Science Foundation of Guangdong Province (2020A1515010548). We acknowledge the allocation time from the SCUTGrid at South China University of Technology.

**Figure legends**

**Figure 1.** The Architecture of FP-GNN. (a) The graph attention network calculates attentions between each node and its neighbors, and then updates the node with those relative attentions. (b) The FP-GNN model combines the information from the molecular graphs and fingerprints to predict molecular properties.

**Figure 2.** Performance of FP-GNN compared to the baseline models on the LIT-PCBA dataset. (a) The NB, SVM, RF, XGBoost and DNN models were established using the Morgan fingerprint as the molecular representation. (b) The NB, SVM, RF, XGBoost and DNN models were established using the mixed fingerprints (MACCS FP, PubChem FP and Pharmacophore ErG FP) as the molecular representation. The performances of models based on the Morgan fingerprint were collected from Jiang et al.[56]. Three graph-based models (GCN, GAT and FP-GNN), as well as models based on the mixed fingerprint were constructed using the same benchmark.

**Figure 3.** Distribution of the ratios of GNN in the FP-GNN models on the 13 public datasets. The details of optimal sets of hypermeters for the 13 public datasets are shown in Supplementary Table 6.

**Figure 4.** Results of the ablation study on the LIT-PCBA dataset.

**Figure 5.** Comparisons of the performance of FP-GNN models based on the Morgan, as well as models based on the mixed of three complementary fingerprints. (a) represents the performance results for three classification datasets (BACE, BBBP and SIDER). (b) represents the performance results for four regression datasets (Lipophilicity, PDBbind-C, PDBbind-F, and PDBbind-R). To ensure the reliability of the results, after optimizing the hyperparameters, the average metric value of the FP-GNN models based on 10 different random seeds was computed as the final result.

**Figure 6.** The anti-noise performances of Attentive FP, HRGCN+, XGBoost and FP-GNN models with different noise rates on the HIV dataset. The anti-noise results for Attentive FP, HRGCN+ and XGBoost models were collected from Wu et al.[54].

**Figure 7.** The importance of molecular structures during the prediction process. The darker the color, the more important are for the structures. Molecules were obtained from the BBBP (the blood-brain barrier penetration) dataset. (a) Molecule 1 is permeable, and the darker colored portion has a higher ClogP, which indicates a stronger lipophilicity. (b) Molecule 2 is impermeable, and the darker colored portion has lower ClogP, which means a weaker lipophilicity. The important portions that were captured by FP-GNN models were consistent with the prediction results.

**Table legends**

**Table 1.** Predictive performance results of FP-GNN on 13 commonly used public datasets.

**Table 2.** Performance of FP-GNN on 14 breast cell line datasets compared to the graph-based DL models.

**Table 3.** The ten most significant bits of the mixed fingerprints on the prediction of the FreeSolv dataset.

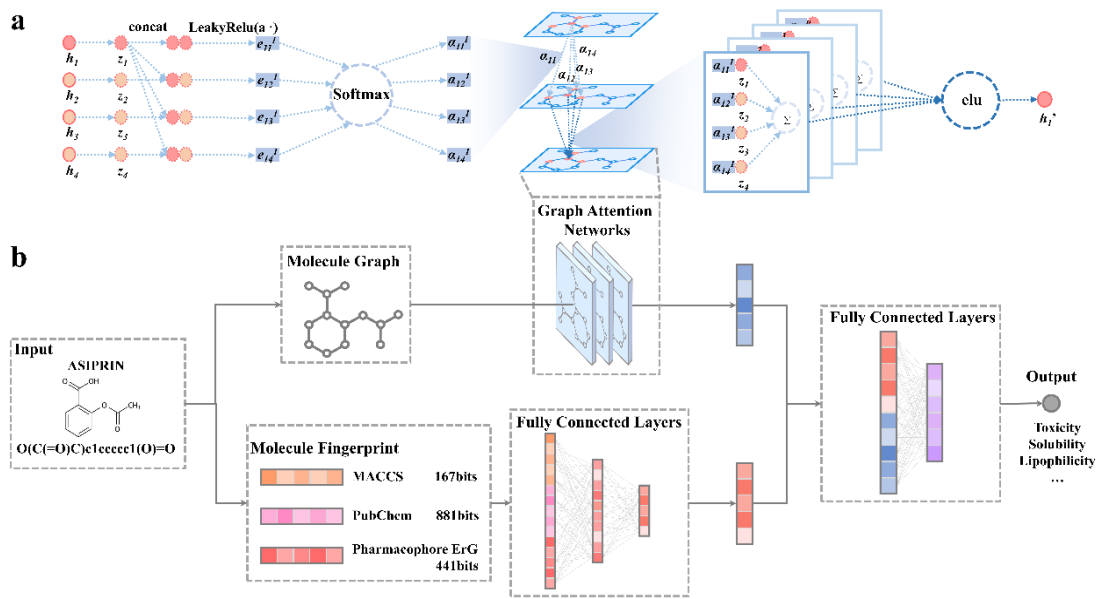

**Figure 1.** The Architecture of FP-GNN. (a) The graph attention network calculates attentions between each node and its neighbors, and then updates the node with those relative attentions. (b) The FP-GNN model combines the information from the molecular graphs and fingerprints to predict molecular properties.

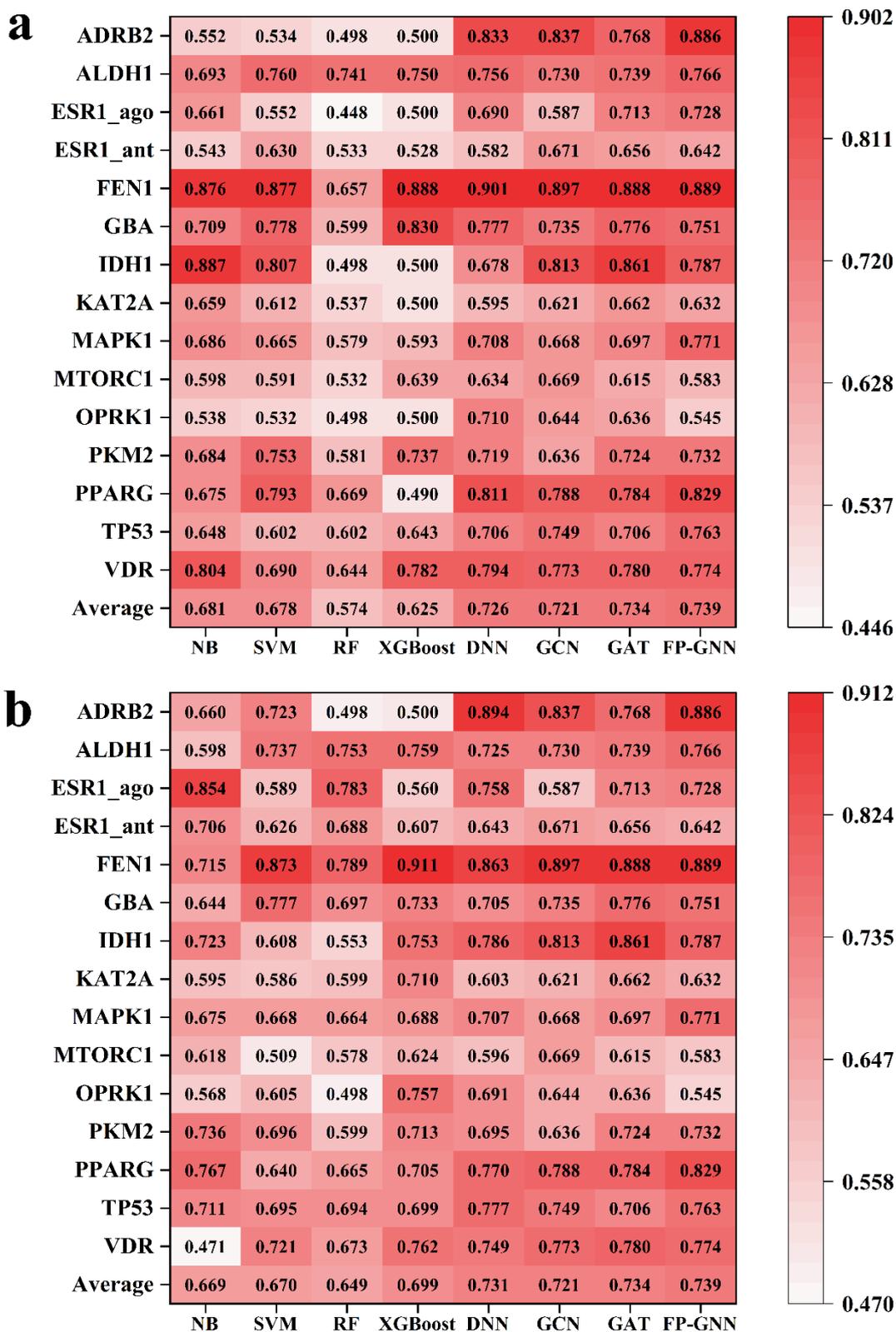

**Figure 2.** Performance of FP-GNN compared to the baseline models on the LIT-PCBA dataset. (a) The NB, SVM, RF, XGBoost and DNN models were established using the Morgan fingerprint as the molecular representation. (b) The NB, SVM, RF, XGBoost and DNN models were established using the mixed fingerprints (MACCS FP, PubChem FP and Pharmacophore ErG FP) as the molecular representation. The performances of models based on the Morgan fingerprint were collected from Jiang et al.[56]. Three graph-based models (GCN, GAT and FP-GNN), as well as models based on the mixed fingerprint were constructed using the same benchmark.

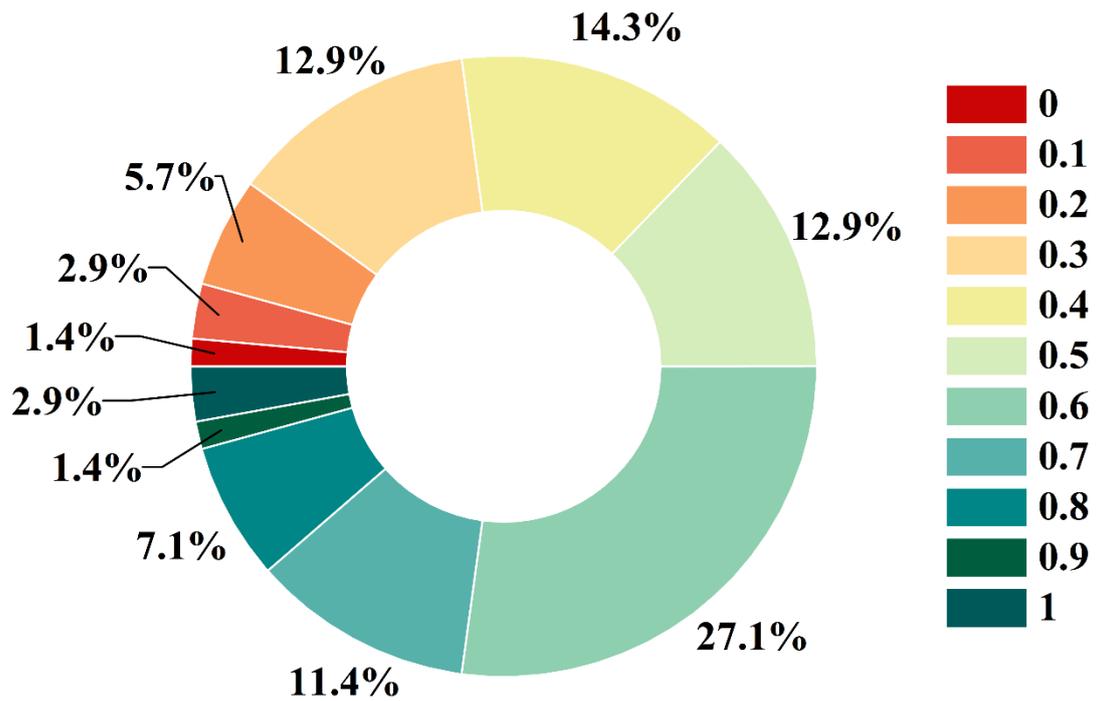

**Figure 3.** Distribution of the ratios of GNN in the FP-GNN models on the 13 public datasets. The details of optimal sets of hypermeters for the 13 public datasets are shown in Supplementary Table 6.

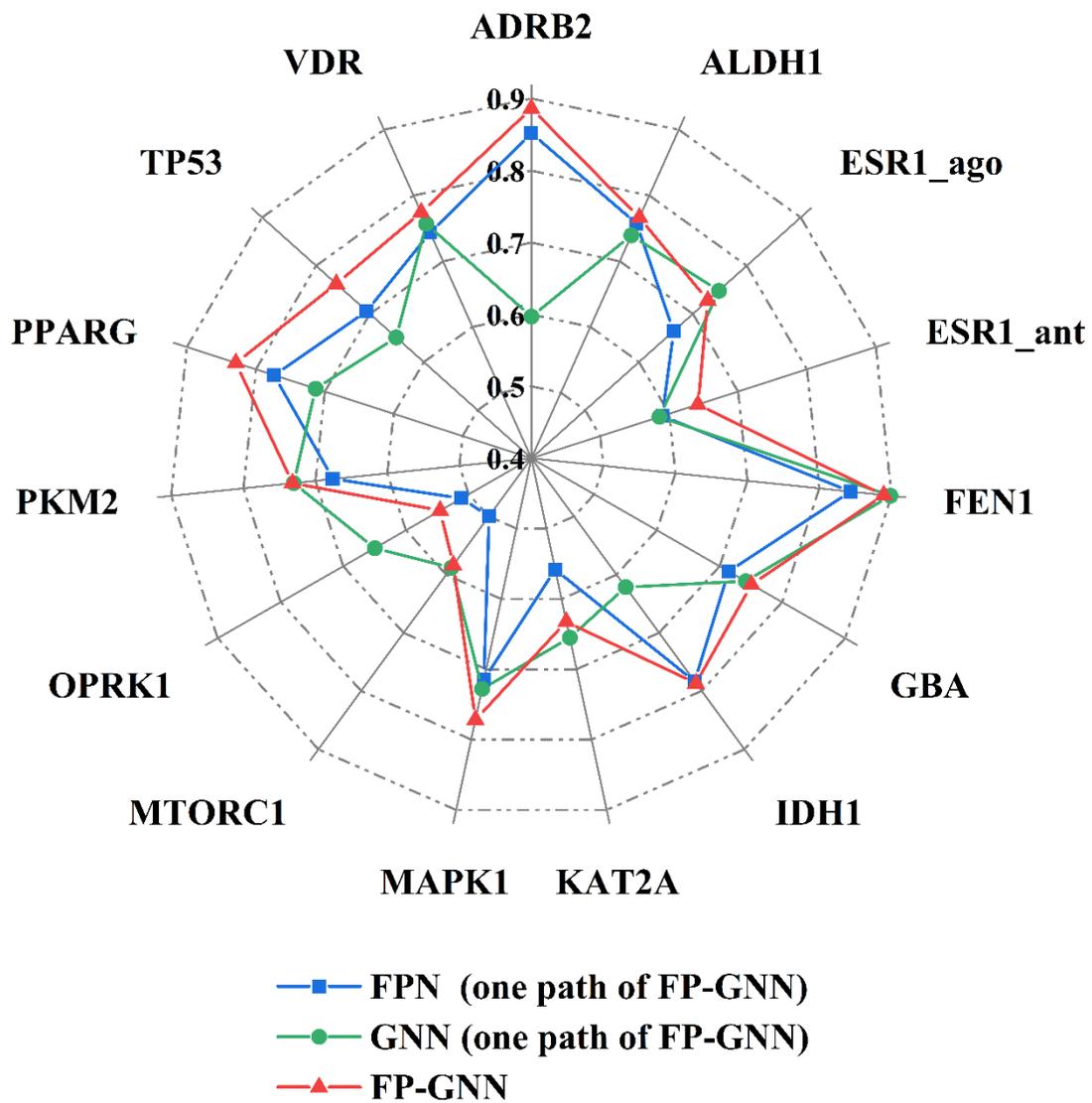

**Figure 4.** Results of the ablation study on the LIT-PCBA dataset.

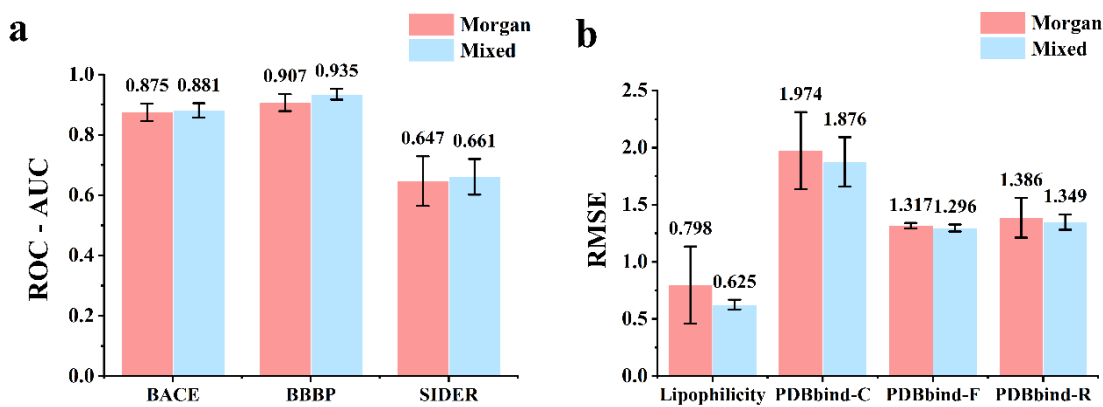

**Figure 5.** Comparisons of the performance of FP-GNN models based on the Morgan, as well as models based on the mixed of three complementary fingerprints. (a) represents the performance results for three classification datasets (BACE, BBBP and SIDER). (b) represents the performance results for four regression datasets (Lipophilicity, PDBbind-C, PDBbind-F, and PDBbind-R). To ensure the reliability of the results, after optimizing the hyperparameters, the average metric value of the FP-GNN models based on 10 different random seeds was computed as the final result.

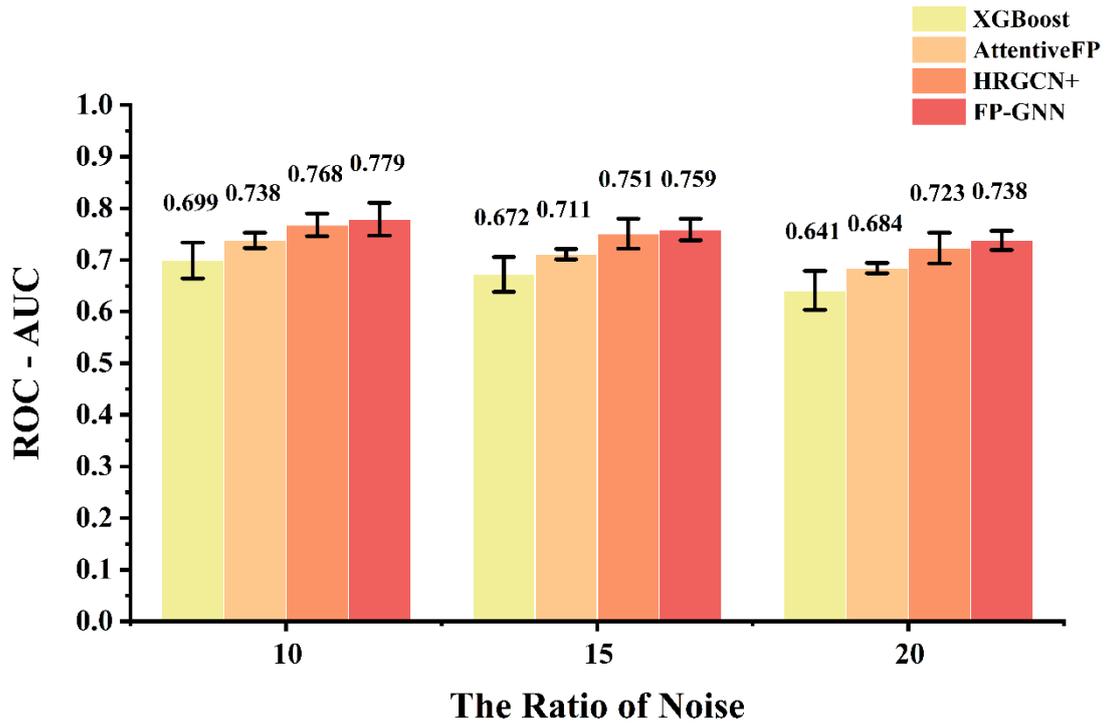

**Figure 6.** The anti-noise performances of Attentive FP, HRGCN+, XGBoost and FP-GNN models with different noise rates on the HIV dataset. The anti-noise results for Attentive FP, HRGCN+ and XGBoost models were collected from Wu et al.[54].

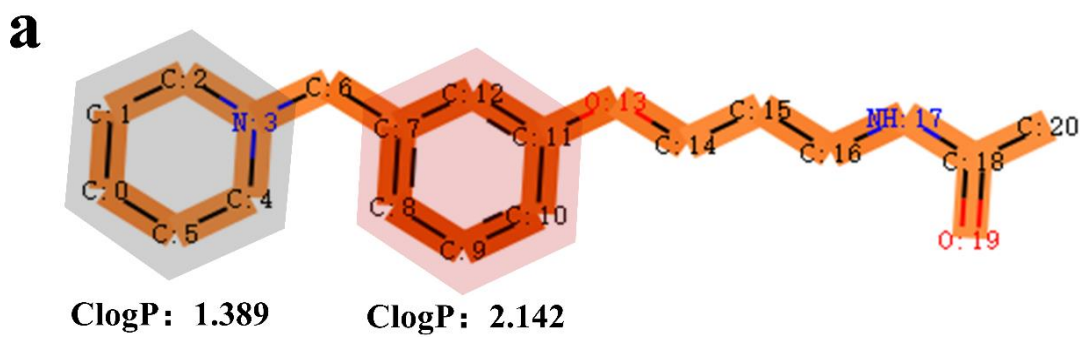

**Molecule 1: permeable**

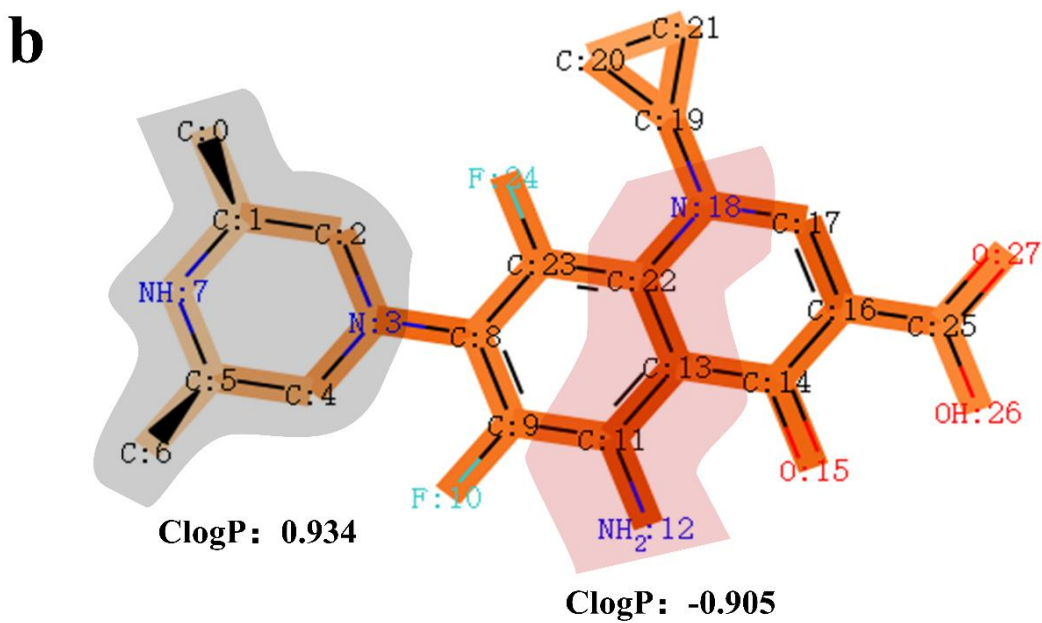

**Molecule 2: impermeable**

**Figure 7.** The importance of molecular structures during the prediction process. The darker the color, the more important are for the structures. Molecules were obtained from the BBBP (the blood-brain barrier penetration) dataset. (a) Molecule 1 is permeable, and the darker colored portion has a higher ClogP, which indicates a stronger lipophilicity. (b) Molecule 2 is impermeable, and the darker colored portion has lower ClogP, which means a weaker lipophilicity. The important portions that were captured by FP-GNN models were consistent with the prediction results.

**Table 1.** Predictive performance results of FP-GNN on 13 commonly used public datasets.

| Dataset | Split type | Metric | MoleculeNet (Graph)[27] | Chemprop (optimized)[20] | Attentive FP[54] | HRGCN+[54] | XGBoost[54] | FP-GNN |
|---|---|---|---|---|---|---|---|---|
| BACE | random | ROC-AUC | | **0.898** | 0.876 | 0.891 | 0.889 | 0.881 |
| | scaffold | ROC-AUC | 0.806 (Weave) | 0.857 | | | | **0.860** |
| HIV | random | ROC-AUC | | **0.827** | 0.822 | 0.824 | 0.816 | 0.825 |
| | scaffold | ROC-AUC | 0.763 (GC) | 0.794 | | | | **0.824** |
| MUV | random | PRC-AUC | **0.109 (Weave)** | 0.053 | 0.038 | 0.082 | 0.068 | 0.090 |
| Tox21 | random | ROC-AUC | 0.829 (GC) | **0.854** | 0.852 | 0.848 | 0.836 | 0.815 |
| BBBP | random | ROC-AUC | | 0.917 | 0.887 | 0.926 | 0.926 | **0.935** |
| | scaffold | ROC-AUC | 0.690 (GC) | 0.886 | | | | **0.916** |
| ClinTox | random | ROC-AUC | 0.832 (Weave) | 0.897 | 0.904 | 0.899 | **0.911** | 0.840 |
| SIDER | random | ROC-AUC | 0.638 (GC) | 0.658 | 0.623 | 0.641 | 0.642 | **0.661** |
| PDBbind-C | random | RMSE | | 1.910 | | | | **1.876** |
| PDBbind-F | random | RMSE | | **1.286** | | | | 1.296 |
| PDBbind-R | random | RMSE | | **1.338** | | | | 1.349 |
| FreeSolv | random | RMSE | 1.150 (MPNN) | 1.009 | 1.091 | 0.926 | 1.025 | **0.905** |
| ESOL | random | RMSE | 0.580 (MPNN) | 0.587 | 0.587 | **0.563** | 0.582 | 0.675 |
| Lipophilicity | random | RMSE | 0.655 (GC) | 0.563 | **0.553** | 0.603 | 0.574 | 0.625 |

Each dataset was split into training, validation and test sets using the corresponding data-split codes from published studies. The FP-GNN models used the same dataset and data split method to fairly compare the MoleculeNet, Chemprop, Attentive FP, HRGCN+ and XGBoost models. Bold font illustrates the models that outperformed all other models. The best graph-based models from MoleculeNet were used, the optimized results of the Chemprop models were from the original study [20], and the best performance results for the Attentive FP, HRGCN+ and XGBoost models were chosen from Wu et al. [54]. MPNN: message passing neural networks; GC: graph convolutional models; and Weave: Weave models.

**Table 2.** Performance of FP-GNN on 14 breast cell line datasets compared to the graph-based DL models.

| Cell Lines | Classification | Compounds | Task metric | Attentive FP[53] | GAT[53] | GCN[53] | MPNN[53] | XGBoost[53] | FP-GNN |
|---|---|---|---|---|---|---|---|---|---|
| MDA-MB-453 | HER-2+ [a] | 440 | ROC-AUC | 0.872 | 0.812 | 0.866 | 0.715 | 0.810 | **0.886** |
| SK-BR-3 | HER-2+ | 2026 | ROC-AUC | 0.805 | 0.840 | 0.839 | 0.760 | 0.848 | **0.852** |
| MDA-MB-435 | HER-2+ | 3030 | ROC-AUC | 0.824 | 0.830 | **0.858** | 0.749 | 0.853 | 0.820 |
| T-47D | Luminal A [b] | 3135 | ROC-AUC | 0.812 | 0.763 | 0.819 | 0.751 | 0.821 | **0.846** |
| MCF-7 | Luminal A | 29378 | ROC-AUC | 0.845 | 0.800 | 0.833 | 0.843 | 0.826 | **0.866** |
| MDA-MB-361 | Luminal B [c] | 367 | ROC-AUC | 0.938 | 0.896 | 0.955 | 0.972 | **0.976** | 0.905 |
| BT-474 | Luminal B | 811 | ROC-AUC | 0.787 | 0.657 | 0.866 | 0.847 | 0.827 | **0.868** |
| BT-20 | TNBC [d] | 292 | ROC-AUC | 0.735 | 0.721 | 0.740 | 0.784 | 0.740 | **0.887** |
| BT-549 | TNBC | 1182 | ROC-AUC | 0.630 | 0.710 | 0.669 | 0.634 | 0.651 | **0.807** |
| HS-578T | TNBC | 469 | ROC-AUC | **0.830** | 0.758 | 0.636 | 0.665 | 0.753 | 0.770 |
| MDA-MB-231 | TNBC | 11202 | ROC-AUC | **0.870** | 0.770 | 0.859 | 0.850 | 0.865 | 0.866 |
| MDA-MB-468 | TNBC | 1986 | ROC-AUC | 0.875 | 0.875 | 0.887 | 0.858 | **0.896** | 0.888 |
| Bcap37 | TNBC | 275 | ROC-AUC | **0.858** | 0.767 | 0.693 | 0.807 | 0.744 | 0.779 |
| HBL-100 | Normal Cell Line | 316 | ROC-AUC | 0.645 | 0.641 | 0.658 | 0.701 | 0.776 | **0.850** |
| | Average | | | 0.809 | 0.774 | 0.798 | 0.781 | 0.813 | **0.849** |

[a]HER-2+: HER2-positive breast cancers. [b]Luminal A: Luminal A breast cancer is hormone-receptor positive (estrogen-receptor and/or progesterone-receptor positive), HER2-negative, with low levels of the protein Ki-67. [c]Luminal B: Luminal B breast cancer is hormone-receptor positive (estrogen-receptor and/or progesterone-receptor positive), HER2 positive or HER2 negative, with high levels of Ki-67. [d]TNBC: triple-negative breast cancer. Each dataset was split into training, validation and test sets using the corresponding data-split codes from He et al. [53]. The FP-GNN models used the same dataset and data split method to fairly compare the Attentive FP, GAT, GCN, MPNN and XGBoost models. Bold font illustrates the models that outperformed all other models. MPNN: message passing neural networks; GCN: graph convolutional networks; GAT: graph attention network.

**Table 3.** The ten most significant bits of the mixed fingerprints on the prediction of the FreeSolv dataset.

| Rank | Importance | Mixed FP Bit | Fingerprint | Fingerprint Bit | Meaning |
|------|------------|--------------|-------------|-----------------|---------|
| 1 | 0.276 | 190 | Pharmacophore ErG | 23 | ('Donor', 'Acceptor', 2) |
| 2 | 0.226 | 189 | Pharmacophore ErG | 22 | ('Donor', 'Acceptor', 1) |
| 3 | 0.196 | 93 | MACCS | 93 | OC(N)C |
| 4 | 0.184 | 1048 | PubChem | 440 | C-C-O=O |
| 5 | 0.177 | 1060 | PubChem | 452 | C-O=O |
| 6 | 0.170 | 111 | MACCS | 111 | NCO |
| 7 | 0.165 | 140 | MACCS | 140 | OH |
| 8 | 0.162 | 1015 | PubChem | 407 | OCP |
| 9 | 0.158 | 276 | Pharmacophore ErG | 109 | ('Donor', 'Aromatic', 4) |
| 10 | 0.156 | 42 | MACCS | 42 | C#N |